\documentstyle[12pt]{article}
\catcode`\@=11

\@addtoreset{equation}{section}
\def\theequation{\arabic{section}.\arabic{equation}}
\catcode`\@=11

\def\section{\@startsection{section}{1}{\z@}{3.5ex plus 1ex minus
   .2ex}{2.3ex plus .2ex}{\large\bf}}

\def\eqnarray{\let\@currentlabel=\theequation\refstepcounter{equation}
    \global\@eqnswtrue
    \global\@eqcnt\z@\tabskip\@centering\let\\=\@eqncr
    $$\halign to \displaywidth\bgroup\@eqnsel\hskip\@centering
      $\displaystyle\tabskip\z@{##}$&\global\@eqcnt\@ne
       \hfil${{}##{}}$\hfil
      &\global\@eqcnt\tw@ $\displaystyle\tabskip\z@{##}$\hfil
       \tabskip\@centering&\llap{##}\tabskip\z@\cr}
\def\lefteqn#1{\hbox to 4\arraycolsep{$\displaystyle #1$\hss}}
\def\thesection{\arabic{section}.}

\def\appendix{\setcounter{section}{0}
        \def\thesection{Appendix.}
        \def\theequation{\Alph{section}.\arabic{equation}}}

\long\def\@makefntext#1{\parindent 0cm\noindent
\hbox to 1em{\hss$^{\@thefnmark}$}#1}
\def\IR{{\hbox{{\rm I}\kern-.2em\hbox{\rm R}}}}
\def\IH{{\hbox{{\rm I}\kern-.2em\hbox{\rm H}}}}
\def\IC{{\ \hbox{{\rm I}\kern-.6em\hbox{\bf C}}}}
\def\IZ{{\hbox{{\rm Z}\kern-.4em\hbox{\rm Z}}}}
\def\rref#1{(\ref{#1})}
\newcommand{\beq}{\begin{equation}}
\newcommand{\eeq}{\end{equation}}
\newcommand{\NPB}[1]{{\sl Nucl.~Phys.}~{\bf B#1}}

\newcommand{\PLB}[1]{{\sl Phys.~Lett.}~{\bf B#1}}

\newcommand{\PRL}[1]{{\sl Phys.~Rev.~Lett.}~{\bf #1}}

\newcommand{\PRD}[1]{{\sl Phys.~Rev.}~{\bf D#1}}

\begin{document}
%%%%%%%%%%%%%%%%%%%%%%%%%%%%%%%%%%%%%%%%%%%%%%%%%%%%%%%%%%%%%%%%%%%%%%%%%%%
%     C I T E . S T Y
%     compressed lists of numerical citations: [11-16]
%     see also OVERCITE.STY and DRFTCITE.STY
%
%     Copyright (C) 1989-1992 by Donald Arseneau
%     These macros may be freely transmitted, reproduced, or modified for
%     non-commercial purposes provided that this notice is left intact.
%
%
%  \@citen contains the code that parses the list of names, ignoring
%  spaces after commas, writes the aux file \citation, and formats the
%  number list.  \citen can be used by itself to give citation numbers
%  without the other formatting; e.g., "See also ref.~\citen{junk}."
%
\def\citen#1{%
\edef\@tempa{\@ignspaftercomma,#1, \@end, }% ignore spaces in parameter list
\edef\@tempa{\expandafter\@ignendcommas\@tempa\@end}%
\if@filesw \immediate \write \@auxout {\string \citation {\@tempa}}\fi
\@tempcntb\m@ne \let\@h@ld\relax \let\@citea\@empty
\@for \@citeb:=\@tempa\do {\@cmpresscites}%
\@h@ld}
%
% for ignoring spaces in the input:
\def\@ignspaftercomma#1, {\ifx\@end#1\@empty\else
   #1,\expandafter\@ignspaftercomma\fi}
\def\@ignendcommas,#1,\@end{#1}
%
% For each citation, check if it is defined, if it is a number, and
% if it is a consecutive number that can be represented like 3-7.
%
\def\@cmpresscites{%
 \expandafter\let \expandafter\@B@citeB \csname b@\@citeb \endcsname
 \ifx\@B@citeB\relax % undefined
    \@h@ld\@citea\@tempcntb\m@ne{\bf ?}%
    \@warning {Citation `\@citeb ' on page \thepage \space undefined}%
 \else%  defined
    \@tempcnta\@tempcntb \advance\@tempcnta\@ne
    \setbox\z@\hbox\bgroup % check if citation is a number:
    \ifnum\z@<0\@B@citeB \relax
       \egroup \@tempcntb\@B@citeB \relax
       \else \egroup \@tempcntb\m@ne \fi
    \ifnum\@tempcnta=\@tempcntb % Number follows previous--hold on to it
       \ifx\@h@ld\relax % first pair of successives
          \edef \@h@ld{\@citea\@B@citeB}%
       \else % compressible list of successives
%         % use \hbox to avoid easy \exhyphenpenalty breaks
          \edef\@h@ld{\hbox{--}\penalty\@highpenalty \@B@citeB}%
       \fi
    \else   %  non-successor--dump what's held and do this one
       \@h@ld \@citea \@B@citeB \let\@h@ld\relax
 \fi\fi%
 \let\@citea\@citepunct
}
%
%%    To put space after the comma, use:
\def\@citepunct{,\penalty\@highpenalty\hskip.13em plus.1em minus.1em}%
%%    For no space after comma, use:
%% \def\@citepunct{,\penalty\@highpenalty}%
%%
%
%  Make \@citex refer to \citen:
%
\def\@citex[#1]#2{\@cite{\citen{#2}}{#1}}%
%
%  Replacement for \@cite.  Give one normal space before the citation,
%  set high penalties for linebreaks,
%
\def\@cite#1#2{\leavevmode\unskip
  \ifnum\lastpenalty=\z@ \penalty\@highpenalty \fi % highpenalty before
  \ [{\multiply\@highpenalty 3 #1% % triple-highpenalties within list
      \if@tempswa,\penalty\@highpenalty\ #2\fi % and before note.
    }]\spacefactor\@m}
\let\nocitecount\relax  % in case \nocitecount was used for drftcite
%
%%%%%%%%%%%%%%%%%%%%%%%%%%%%%%%%%%%%%%%%%%%%%%%%%%%%%%%%%%%%%%%%%%%%%%%%%%
\begin{titlepage}
\vspace{.5in}
\begin{flushright}
UCD-94-32\\
NI-94011\\
gr-qc/9409052\\
September 1994\\
\end{flushright}
\vspace{.5in}
\begin{center}
{\Large\bf
 The Statistical Mechanics\\[1ex]
 of the (2+1)-Dimensional Black Hole}\\
\vspace{.4in}
{S.~C{\sc arlip}\footnote{\it email: carlip@dirac.ucdavis.edu}\\
       {\small\it Department of Physics}\\
       {\small\it University of California}\\
       {\small\it Davis, CA 95616}\\{\small\it USA}}
\end{center}

\vspace{.5in}
\begin{center}
\begin{minipage}{4in}
\begin{center}
{\large\bf Abstract}
\end{center}
{\small
The presence of a horizon breaks the gauge invariance of
(2+1)-dimensional general relativity, leading to the appearance
of new physical states at the horizon.  I show that the entropy
of the (2+1)-dimensional black hole can be obtained as the
logarithm of the number of these microscopic states.}
\end{minipage}
\end{center}
\end{titlepage}
\addtocounter{footnote}{-1}

\section{Introduction}

Black holes possess a temperature and an entropy, and obey the usual
laws of thermodynamics.  Despite twenty years of research, however,
black hole thermodynamics remains something of an anomaly.  The
thermal properties of ordinary physical systems arise from the
statistical mechanics of microscopic states.  But a classical black
hole is completely characterized by its mass and angular momentum,
leaving little room for additional microscopic physics.  Black holes
have entropy, but we do not know why.

The dependence of black hole temperature and entropy on Planck's
constant suggests that these quantities are fundamentally quantum
mechanical in nature.  The goal of this paper is to demonstrate a
mechanism whereby black hole entropy---at least for the (2+1)-dimensional
black hole of Ba\~nados, Teitelboim, and Zanelli \cite{BTZ}---can be
obtained by counting quantum gravitational states at the horizon.
The restriction to 2+1 dimensions is, of course, a serious limitation,
but if enough new states can be found to account for black hole entropy
in this simple setting, it is reasonable to hope that the same will
happen in the vastly richer arena of realistic (3+1)-dimensional
gravity.

The basic argument is quite simple.  Begin by considering general
relativity on a manifold $M$ with boundary.  We ordinarily split the
metric into true physical excitations and ``pure gauge'' degrees of
freedom that can be removed by diffeomorphisms of $M$.  But the presence
of a boundary alters the gauge invariance of general relativity: the
infinitesimal transformations $g\rightarrow g + {\cal L}_\xi g$ must
now be restricted to those generated by vector fields $\xi$ with no
component normal to the boundary, that is, true diffeomorphisms that
preserve $\partial M$.  As a consequence, some degrees of freedom
that would naively be viewed as ``pure gauge'' become dynamical,
introducing new degrees of freedom associated with the boundary.

Now, the event horizon of a black hole is not a true boundary, although
the black hole complementarity approach of Susskind et al.\ \cite{Suss}
suggests that it might be appropriately treated as such.  Regardless
of one's view of that program, however, it is clear that in order to
ask quantum mechanical questions about the behavior of black holes, one
must put in ``boundary conditions'' that ensure that a black hole is
present.  This means requiring the existence of a hypersurface with
particular metric properties---say, those of an apparent horizon.

The simplest way to do quantum mechanics in the presence of such a
surface is to quantize fields separately on each side, imposing the
appropriate correlations as boundary conditions.  In a path integral
approach, for instance, one can integrate over fields on each side,
equate the boundary values, and finally integrate over those boundary
values compatible with the existence of a black hole.  But this
process again introduces boundary terms that restrict the gauge
invariance of the theory, leading once more to the appearance of
new degrees of freedom at the horizon that would otherwise be
treated as unphysical.

My suggestion is that black hole entropy is determined by counting these
would-be gauge degrees of freedom.  The resulting picture is similar
to Maggiore's membrane model of the black hole horizon \cite{Mag}, but
with a particular derivation and interpretation of the ``membrane''
degrees of freedom.

The analysis of this phenomenon is fairly simple in 2+1 dimensions.
It is well known that (2+1)-dimensional gravity can be written as a
Chern-Simons theory \cite{Acha,Witten}, and it is also a standard
result that a Chern-Simons theory on a manifold with boundary
induces a dynamical Wess-Zumino-Witten (WZW) theory on the boundary
\cite{zoo,EMSS}.  In the presence of a cosmological constant $\Lambda
= -1/\ell^2$ appropriate for the (2+1)-dimensional black hole, one
obtains a slightly modified $\hbox{SO}(2,1)\!\times\!\hbox{SO}(2,1)$
WZW model, with coupling constant
\beq
k = {\ell\sqrt{2}\over 8G} .
\label{a1}
\eeq
This model is not completely understood, but in the large $k$---i.e.,
small $\Lambda$---limit, it may be approximated by a theory of six
independent bosonic oscillators.  I show below that the Virasoro
operator $L_0$ for this theory takes the form
\beq
L_0 \sim N - \left({r_+\over4G}\right)^2 ,
\label{a2}
\eeq
where $N = \sum_{i=1}^6 N_i$ is a number operator and $r_+$ is the
horizon radius.  It is a standard result of string theory \cite{Wein}
that the number of states of such a system behaves asymptotically as
\beq
n(N) \sim \exp\left\{\pi\sqrt{6\cdot{2N\over3}}\right\} .
\label{a3}
\eeq
If we demand that $L_0$ vanish---physically, requiring states to
be independent of the choice of origin of the angular coordinate at the
horizon---we thus obtain
\beq
\log n(r_+) \sim {2\pi r_+\over 4G} ,
\label{a4}
\eeq
precisely the right expression for the entropy of the (2+1)-dimensional
black hole \cite{BTZ,myBH}.

I now turn to the details of this analysis.

\section{Chern-Simons Theory, WZW Models, and Gravity}

The most thoroughly studied example of boundary dynamics arising from a
simple ``bulk'' action is that of Chern-Simons theory.  The Chern-Simons
action for a three-manifold $M$ is
\beq
I_{\hbox{\scriptsize CS}} = {k\over4\pi}\int_M
  \hbox{Tr}\left( A\wedge dA + {2\over3}A\wedge A\wedge A \right) ,
\label{b1}
\eeq
where $A = A^a_\mu T^{\hphantom 1}_a dx^\mu$ is a gauge field
(connection one-form) for a group $G$ whose Lie algebra is generated
by $\{T_a\}$.  If $M$ is closed, this action is invariant under gauge
transformations
\beq
\bar A \rightarrow A = g^{-1}dg + g^{-1}\bar A g  .
\label{b2}
\eeq
If $M$ has a boundary, however, this invariance is broken: the integrand
of \rref{b1} is invariant only up to a total divergence, which can lead
to a  nontrivial term on $\partial M$.  Moreover, in order for the
theory to admit classical solutions, $I_{\hbox{\scriptsize CS}}$ must
be supplemented by a surface term, whose exact form depends on the choice
of boundary conditions.  In particular, if $\partial M$ has the topology
of a cylinder, parametrized by an angular coordinate $\phi$ and a linear
coordinate $v$, and if $A_\phi$ is fixed at the boundary, then the
required term is
\beq
I^\prime_{\hbox{\scriptsize CS}}
  = -{k\over4\pi}\int_{\partial M} \hbox{Tr}A_\phi A_v ,
\label{b3}
\eeq
which further breaks the gauge invariance of the ``bulk'' action.

To isolate the gauge dependence of the action, it is useful to partition
the space of connections into gauge orbits parametrized by gauge-fixed
connections $\bar A$, with points on each orbit labeled by group elements
$g$.  In the path integral context, this may be recognized as the first
step in Faddeev-Popov gauge-fixing.  Writing a general connection in the
form \rref{b2}, it is easy to show that \cite{Ogura,myLiou}
\beq
(I_{\hbox{\scriptsize CS}}+I^\prime_{\hbox{\scriptsize CS}})[A] =
  (I_{\hbox{\scriptsize CS}}+I^\prime_{\hbox{\scriptsize CS}})[\bar A]
  - kI^{\raise2pt\hbox{$\scriptstyle +$}}_{\hbox{\scriptsize WZW}}[g,\bar A],
\label{b4}
\eeq
where
\begin{eqnarray}
I^{\raise2pt\hbox{$\scriptstyle +$}}_{\hbox{\scriptsize WZW}}[g,\bar A]
  &=& {1\over4\pi}\int_{\partial M}\hbox{Tr}
  \left(g^{-1}\partial_\phi g\right)\left(g^{-1}\partial_v g\right)\
  \nonumber\\
  &\ & +\,{1\over2\pi}\int_{\partial M}\hbox{Tr}
  \left(g^{-1}\partial_v g\right)\left(g^{-1}\bar A_\phi g\right)
  + {1\over12\pi}\int_M\hbox{Tr}\left(g^{-1}dg\right)^{\lower2pt
  \hbox{$\scriptstyle 3$}}
\label{b5}
\end{eqnarray}
is the standard chiral Wess-Zumino-Witten action on $\partial M$.  This
action can be obtained in a number of ways \cite{EMSS}, but the derivation
given here points to a particular physical interpretation: the boundary
variables are would-be ``pure gauge'' degrees of freedom that become
dynamical because the presence of a boundary restricts the allowed gauge
transformations.

As first shown by Ach\'ucarro and Townsend \cite{Acha}, (2+1)-dimensional
gravity can itself be written as a Chern-Simons theory.  In particular,
for the case of a cosmological constant $\Lambda=-1/\ell^2$, one can
define two $\hbox{SO}(2,1)$ gauge fields
\beq
A^a = \omega^a + {1\over\ell}e^a , \quad
  \tilde A = \omega^a - {1\over\ell}e^a ,
\label{b6}
\eeq
where $e^a = e^a_\mu dx^\mu$ is a triad and $\omega^a = {1\over2}
\epsilon^{abc}\omega_{\mu bc}dx^\mu$ is a spin connection.  The
first-order form of the Einstein action is then
\beq
I_{\hbox{\scriptsize grav}}
  = I_{\hbox{\scriptsize CS}}[A] - I_{\hbox{\scriptsize CS}}[\tilde A] .
\label{b7}
\eeq
The value of the coupling constant $k$ depends on normalization.  For
this paper, I choose
\beq
(T_a)_b{}^c = -\epsilon_{abd}\eta^{dc}, \quad
  \eta_{ab} = \hbox{diag}\,(-1,1,1), \quad \epsilon_{012}=1 ,
\label{b7a}
\eeq
and define $\hbox{Tr}$ as the ordinary matrix trace, so
\begin{eqnarray}
&&[T_a,T_b] = f_{ab}{}^c T_c = \epsilon_{abd}\eta^{dc}T_c , \quad
  \hbox{Tr}\,T_aT_b = \hat g_{ab} = 2\eta_{ab} , \quad
  g_{\mu\nu}=\hat g_{ab}e^a_\mu e^b_\nu=2\eta_{ab}e^a_\mu e^b_\nu ;
  \nonumber\\
&&\hat g^{ad}\hat g^{be}f_{ab}{}^cf_{de}{}^f = Q\hat g^{cf} \ \
 \hbox{with $Q=-1$} .
\label{b8}
\end{eqnarray}
Then $k$ is given by \rref{a1}, as is most easily checked by comparing
the value of the action \rref{b7} at a classical solution to the
corresponding Einstein action.  The diffeomorphism invariance of
general relativity translates into gauge invariance of the Chern-Simons
action: appropriate combinations of gauge transformations of $A$ and
$\tilde A$ are equivalent to diffeomorphisms \cite{Witten}.  We should
thus expect (2+1)-dimensional gravity to induce a pair of $\hbox{SO}(2,1)$
WZW actions on $\partial M$, whose degrees of freedom correspond in some
sense to deformations of the horizon.

\section{The Boundary Action}

To determine the exact form of these boundary WZW actions, we must
now choose boundary conditions for $A$ and $\tilde A$.  Ideally, we
would require the existence of a black hole by imposing the requirement
that an event horizon be present.  Unfortunately, a genuine event horizon
is a complicated global object, and it is difficult to translate its
existence into local boundary conditions.  Let us therefore impose
the simpler requirement that $\partial M$ be an apparent horizon
(strictly speaking, a trapping horizon \cite{Hayward}).

To do so, we write the metric in double null (light cone) coordinates
\cite{Hayward},
\beq
ds^2 = -2 e^{-f} dudv + r^2(u,v)\left( d\phi + N^u du + N^v dv\right)^2 .
\label{c1}
\eeq
It may then be shown that the spin connection $\omega^+ = \omega^0 +
\omega^1$ is
\beq
\omega^+ = 2re^{f/2}\,\theta^+ \left( d\phi + N^u du + N^v dv\right)
   + B^+ du ,
\label{c2}
\eeq
where $\theta^+(u,v,\phi)$ is the expansion of the outgoing null
geodesic congruence at $(u,v,\phi)$ and $B^+$ is a complicated but
irrelevant function.  For a circle $(u_0,v_0,\phi)$ to lie on an
apparent horizon, we require that $\theta^+(u_0,v_0,\phi)$ vanish.
If in addition the stress-energy tensor vanishes at $(u_0,v_0,
\phi)$---or, more narrowly, if $T^{++}$ vanishes---then it is easy
to show (classically) that the apparent horizon is null at that point.
{}From \rref{c2}, the boundary conditions are thus $\omega^+_\phi =
\omega^+_v = 0$, or
\beq
A^+_\phi = A^+_v = \tilde A^+_\phi = \tilde A^+_v = 0 .
\label{c3}
\eeq

These conditions are not quite sufficient; we must also prescribe
appropriate boundary values for $A^2$ and $\tilde A^2$ at $\partial M$.
As shown in \cite{myBH}, the right choice for black hole thermodynamics
is to fix the horizon radius $r_+$ and either a component $p_+$ of the
extrinsic curvature or its canonical conjugate, the shift vector $N^\phi$.
In the coordinates \rref{c1}, $p_+$ is determined by the spin connection
component $\omega^2_\phi$, and suitable boundary conditions are
\beq
e^2_\phi = {r_+\over\sqrt{2}} , \quad \omega^2_\phi = \bar\omega .
\label{c4}
\eeq
(The factor of $\sqrt2$ comes from the normalizations \rref{b8}.)

The boundary conditions \rref{c3}--\rref{c4} are not, of course,
diffeomorphism-invariant.  This is as it should be, since our aim is
to separate out the ``diffeomorphism'' degrees of freedom at $\partial M$.
The boundary conditions are, however, invariant under rigid translations
$\phi\rightarrow\phi+c(v)$, that is, time-dependent shifts of the
origin of $\phi$.  Thus while most of the diffeomorphisms of the horizon
will be absorbed into the new dynamical fields of the WZW model at
$\partial M$, these rigid translations remain as symmetries.

Now, the physical meaning of $r_+$ is clear---$2\pi r_+$ is the
circumference of the boundary $\partial M$, on or off shell.  The
interpretation of $\bar\omega$ is more problematic.  On shell, it may
be shown that in Kruskal coordinates, $\bar\omega = 2r_-/\ell$, where
$r_-$ is the value of $r$ at the inner horizon.  But this relationship
is coordinate-dependent, and its off-shell generalization is not at
all obvious; for an arbitrary metric, $\bar\omega$ depends not only
on the horizon geometry, but also on normal derivatives.  Moreover,
even on shell, $\bar\omega$ is determined only modulo an integer in
the Euclideanized theory \cite{myBH}.

It would therefore be preferable to fix the shift vector $N^\phi$---or
equivalently, $e^2_v$---at $\partial M$.  Unfortunately, this would
lead to boundary conditions that mix $A$ and $\tilde A$, making the
induced boundary action much more complicated.  I shall instead argue
as follows.  A standard choice of boundary conditions for the black
hole is to set $N^\phi=0$ at $\partial M$, but this choice is somewhat
conventional, since $N^\phi$ is determined only up to an integration
constant \cite{BTZ}, which may be shifted by the rigid rotations
of the horizon discussed above.  So let us instead require that
$\partial_\phi N^\phi = 0$, and sum over the constant values of the
shift vector to count macroscopically indistinguishable states.  Since
$N^\phi$ and $\bar\omega$ are canonically conjugate, this should be
equivalent to summing over constant values of $\bar\omega$.  I will
therefore adopt the boundary conditions \rref{c3}--\rref{c4}, and
integrate over $\bar\omega$ to count states.

Given the boundary conditions \rref{c3}--\rref{c4}, the induced action on
$\partial M$ is not hard to determine.  One obtains
\beq
I[g,\bar A] =
  - kI^{\raise2pt\hbox{$\scriptstyle +$}}_{\hbox{\scriptsize WZW}}[g,\bar A]
  - {k\over2\pi}\int_{\partial M} \left(g^{-1}\partial_v g + g^{-1}A_v g
  \right)^+ \left(g^{-1}\partial_\phi g + g^{-1}\bar A_\phi g\right)^-
\label{c5}
\eeq
with a similar expression for $\tilde A$.  The last term in \rref{c5}
is most easily understood by noticing that the Chern-Simons boundary
action \rref{b3} is appropriate for fixing $A^+_\phi$ and $A^-_\phi$;
if we wish instead to fix $A^+_\phi$ and $A^+_v$, we need an additional
boundary term of the form $\int\!A^-_\phi A^+_v$.  Note that \rref{c5} is
not a ``gauged WZW action'' in the usual sense of the term, since the
fields $\bar A$ are fixed by the boundary data, and not integrated out.

The action $I[g,\bar A]$ is no longer quite the standard $\hbox{SO}(2,1)$
WZW action, but it is classically equivalent.  Indeed, if we define an
element $h$ of $\hbox{SO}(2,1)$ by
\beq
\partial_v h\cdot h^{-1}=(g^{-1}\partial_v g+g^{-1}\bar A_v g)^+\, T^- ,
\label{c6}
\eeq
it follows from the Polyakov-Wiegmann formula \cite{PW} that
\beq
I[g,\bar A] =
-kI^{\raise2pt\hbox{$\scriptstyle +$}}_{\hbox{\scriptsize WZW}}[gh,\bar A] .
\label{c7}
\eeq
Conversely, $h$ is determined from $gh$ by the condition
\beq
\left( h \cdot J_v[gh] \cdot h^{-1} \right)^+ = 0 ,
\label{c8}
\eeq
where $J[g] = g^{-1}\partial_v g + g^{-1}\bar A_v g$.

In the quantum theory, the change of variables from $g$ to $gh$ will
lead to a Jacobian, which can be most easily determined by using the
Gauss decomposition of $g$,
\beq
g = \left(\begin{array}{cc} 1&a\\0&1\end{array}\right)
    \left(\begin{array}{cc} e^\lambda&0\\0&e^{-\lambda}\end{array}\right)
    \left(\begin{array}{cc} 1&0\\b&1\end{array}\right) , \qquad
h = \left(\begin{array}{cc} 1&0\\\hat h&1\end{array}\right) .
\label{c9}
\eeq
In this representation, the change of variables from $b$ to $b+\hat h$
is relatively easy to evaluate; one finds a Jacobian of the form
\beq
{\cal J} = \left|\hbox{det}\left(\partial_v^{-1}A^2_v\right)\right|
\label{c10}
\eeq
with $A^2_v$ defined by \rref{b2}.  This Jacobian will renormalize terms
in the action \rref{c5}, and its careful treatment is necessary for
a full evaluation of the boundary WZW theory.  (A similar deformation
has been considered by F\"orste \cite{For}.)   In the large $k$, or
semiclassical, limit, however, it should be possible to neglect this
correction.  For the purposes of this paper, I will therefore work with
the usual $\hbox{SO}(2,1)\!\times\!\hbox{SO}(2,1)$ WZW action.

\section{Counting States}

It remains for us to count the states of this induced boundary theory.
Observe first that a WZW model is completely characterized by a current
algebra \cite{GepWit,KZ,Papa}
\beq
[J^a_m,J^b_n] = if^{ab}{}_cJ^c_{m+n} - km{\hat g}^{ab}\delta_{m+n,0} ,
  \quad [\tilde J^a_m,\tilde J^b_n] = if^{ab}{}_c\tilde J^c_{m+n}
  + km{\hat g}^{ab}\delta_{m+n,0} ,
\label{d1}
\eeq
where the currents $J^a_n$ and $\tilde J^a_n$ are defined by the
expansions \cite{Ban}
\beq
A^a_\phi=-{1\over k}\sum_{n=-\infty}^\infty J^a_n e^{in\phi} ,\quad
\tilde A^a_\phi={1\over k}\sum_{n=-\infty}^\infty\tilde J^a_n e^{in\phi} .
\label{d2}
\eeq
Here, $A^a$ and ${\tilde A^a}$ are the gauge fields of equation
\rref{b2}---insertion of that equation into \rref{d2} gives the usual
dependence of the currents on $g$ and $\tilde g$, with signs determined
by \rref{b7} and \rref{c7}---and indices are raised and lowered with
the metric $\hat g_{ab}$ of \rref{b8}.  To find the Hilbert space, we
must thus find an appropriate representation of the affine Lie algebra
\rref{d1}.

The standard choice is a highest weight representation.  We start with
a vacuum multiplet $|\Omega\rangle$ that is annihilated by the $J^a_n$ and
$\tilde J^a_n$ with $n\!>\!0$, and that transforms under a representation
of the $\hbox{SO}(2,1)\!\times\!\hbox{SO}(2,1)$ generated by the zero-modes
$J^a_0$ and $\tilde J^a_0$.  This representation is determined by the
boundary conditions \rref{c3}--\rref{c4}, which imply that
\beq
(J_0)^2\,|\Omega\rangle
  = 2k^2\left(\bar\omega +
   {r_+\over\sqrt{2}\ell}\right)^{\lower2pt\hbox{$\scriptstyle2$}}
   |\Omega\rangle, \quad
(\tilde J_0)^2\,|\Omega\rangle
  = 2k^2\left(\bar\omega
   - {r_+\over\sqrt{2}\ell}\right)^{\lower2pt\hbox{$\scriptstyle2$}}
   |\Omega\rangle .
\label{d2a}
\eeq
Note that the positivity of $(J_0)^2$ and $(\tilde J_0)^2$ implies that
the relevant representation must be in the continuous series. (See
\cite{Dixon} for a nice discussion of $\hbox{SO}(2,1)$ representations.)
The zero-modes can alternatively be obtained heuristically from section
2.3 of reference \cite{EMSS}, treating the term $\bar A_\phi$ in \rref{b5}
as a ``source'' term, which determines the relevant representation of
the affine algebra.  Equivalently, in the coadjoint orbit approach to
quantization \cite{Alek}, the $\bar A$ term determines the appropriate
orbit in $\hbox{SO}(2,1)\!\times\!\hbox{SO}(2,1)$.

The remaining states are now obtained by acting on $|\Omega\rangle$
with raising operators $J^a_{-n}$ and $\tilde J^a_{-n}$.  Because
of the commutators \rref{d1}, this process is rather nontrivial, and
the Hilbert space of the $\hbox{SO}(2,1)$ WZW model is not fully
understood (but see \cite{Dixon,Hwang,Hwang2,Ram,Imb}).  In the large
$k$ limit, however, the components of the currents $J$ and $\tilde J$
decouple, and the states can be approximated as those of a six-dimensional
bosonic string theory.  This phenomenon is most easily seen by rescaling
the currents in \rref{d1} by $k^{-1/2}$; in the large $k$ limit, the
terms involving the structure constants drop out, leaving a set of
$\widehat{u(1)}$ commutators at level $\pm 1$.

In string theory, one must impose the added restriction that physical
states be annihilated by the Virasoro generators $L_n$ ($n>0$).  This
condition comes from the requirement of diffeomorphism invariance;
classically, the $L_n$ generate the diffeomorphisms of the circle.
As we saw in the last section, this is not an appropriate requirement
here: the boundary conditions \rref{c3}--\rref{c4} are not
diffeomorphism-invariant, and indeed, the existence of our boundary
degrees of freedom directly reflect this noninvariance.  Our boundary
conditions are, however, still invariant under time-dependent rigid
translations of $\phi$, and we must therefore require the corresponding
invariance of the states, i.e., the vanishing\footnote{Normal ordering
introduces an ambiguity in $L_0$, and in string theory the appropriate
condition is that $L_0=1$.  I do not know whether a similar adjustment
is needed in (2+1)-dimensional gravity, but a small normal-ordering
constant will not qualitatively affect the conclusions of this paper.}
of $L_0$.  The condition $L_0|\psi\rangle=0$ may be viewed as a last
remnant of the Wheeler-DeWitt equation.

In the conventions of this paper, the Virasoro operator $L_0$ for the
action \rref{b7} is
\beq
L_0 = - {1\over 2k-Q}\sum_{n=-\infty}^\infty
      \colon J^a_{-n}J^b_n\colon \hat g_{ab}
      + {1\over 2k+Q}\sum_{n=-\infty}^\infty
      \colon \tilde J^a_{-n}\tilde J^b_n\colon \hat g_{ab} ,
\label{d3}
\eeq
which satisfies
\beq
[L_0, J^a_n] = -n J^a_n ,\ \quad [L_0, \tilde J^a_n] = -n \tilde J^a_n .
\label{d4}
\eeq
These commutation relations imply that the non-zero mode contributions
to $L_0$ take the form of number operators, which in the large $k$
limit can be taken to be independent, while the zero-mode contributions
are determined by \rref{d2a}.  Combining terms and using the normalizations
\rref{b8}, we find
\beq
L_0 = \sum_{i=1}^6 N_i
  + {4k^2\over4k^2-1}\left(
    \bar\omega - {\sqrt{2}kr_+\over\ell}\right)^2
  - {2k^2r_+^2\over\ell^2}
\label{d5}
\eeq
with $k=\ell\sqrt2/8G$.  The condition $L_0|\psi\rangle=0$ thus determines
$\sum N_i$ in terms of $r_+$ and $\bar\omega$.

Now, given a set of independent number operators $N_i$, it is fairly
easy to determine the number of states.  A fixed component $J$ of the
current creates states of the form
\beq
|(n_1,a_1),(n_2,a_2),\dots\rangle =
(J_{-n_1})^{a_1} (J_{-n_2})^{a_2} \dots |\Omega\rangle ,
\label{d5a}
\eeq
for which $N = \sum a_in_i$.  The number of states is then given by
the number of ways of writing $N$ in this form.  This is essentially
the partition function of number theory \cite{part}, whose asymptotic
behavior is given by \rref{a3}; the factor of $6$ in the square root
is the number of independent $N_i$.  A similar expression occurs for
unitary representations of arbitrary affine Lie algebras based on
compact groups \cite{Kac}, and has been generalized to at least some
representations of affine $\hbox{SO}(2,1)$ \cite{Wak}.  The asymptotic
behavior can also be derived (for the discrete series) from the character
formulas of Henningson et al.\ \cite{Hwang} and Dixon and Lykken
\cite{Dixon2}, which can be rewritten to give a generating function
for $n(N)$ in terms of theta functions.

As argued in the last section, we should now integrate over $\bar\omega$
to obtain the total number of macroscopically indistinguishable states.
With $n(N)$ given by \rref{a3} and $N$ determined by \rref{d5}, the
dominant contribution will come from
$$\bar\omega \sim {\sqrt{2}kr_+\over\ell} ,$$
and the total number of states will have the asymptotic behavior
\rref{a4}, as claimed.  We have thus found a good candidate for a
set of microscopic states whose statistical mechanics could explain
the entropy of the (2+1)-dimensional black hole.

Two cautionary remarks are necessary regarding the representation of the
current algebra \rref{d1} used here.  First, the analysis has thus far
ignored the degeneracy of the vacuum $|\Omega\rangle$.  All unitary
irreducible representations of $\hbox{SO}(2,1)\!\times\!\hbox{SO}(2,1)$
are infinite-dimensional---one may act on a highest weight state
$|\Omega\rangle$ with an arbitrary number of factors $(J_0^1-iJ_0^2)$
and $(\tilde J_0^1-i\tilde J_0^2)$ to obtain new states with the same
value of $L_0$.  The states \rref{d5a} have the same degeneracy, and
$n(N)$ actually counts global $\hbox{SO}(2,1)\!\times\!\hbox{SO}(2,1)$
representations rather than individual vectors in each representation.
Second, the representations considered here---like all highest weight
representations of affine $\hbox{SO}(2,1)$ \cite{Chari}---are nonunitary:
states such as $J^0_{-N}|\Omega\rangle$ are easily seen to have negative
norm.

It is not clear whether either of these issues presents a serious obstacle
for a thermodynamic interpretation of our new boundary states.  Nonunitary
conformal field theories occur elsewhere in physics \cite{Cardy,Cohn};
since our boundary states are not yet interacting with any external fields,
the appearance of negative-norm states need not be fatal.  Indeed, one
expects the (2+1)-dimensional black hole coupled to matter to be unstable
against Hawking radiation, and the appearance of negative-norm states in
the uncoupled theory may be seen as a sign of this instability.  It would
be interesting to find a suitable ``Euclidean'' continuation of this model
in which the gauge group $\hbox{SO}(2,1)\!\times\!\hbox{SO}(2,1)$ is
replaced by $\hbox{SO}(3)\!\times\!\hbox{SO}(3)$; it is likely that such
a substitution would simultaneously provide a unitary Hilbert space and
remove the infinite vacuum degeneracy.

\section{Next Steps}

The results of this work strongly suggest that black hole entropy has
a natural microscopic, ``statistical mechanical'' origin.  A number of
important questions remain, however, both in the (2+1)-dimensional
model and in 3+1 dimensions.

In 2+1 dimensions, it is important to understand the boundary WZW model
in more detail.  In particular, the effect of the Jacobian \rref{c10}
needs further investigation, as does the physical significance of the
boundary variable $\bar\omega$.  $\hbox{SO}(2,1)$ WZW models are not yet
well understood---in particular, the proper choice of representation
of the affine $\hbox{SO}(2,1)$ algebra, discussed at the end of the
last section, is not clear---but a good deal of research on this
subject is now in progress.  Ultimately, of course, a thermodynamic
interpretation will require coupling the horizon degrees of freedom to
external fields.  This is a difficult problem, since matter couplings
remove much of the simplicity of (2+1)-dimensional gravity.

I do not know whether the picture presented here will help to resolve
the ``information loss'' paradox of black hole physics.  It is worth
noting that $N\!\sim\! k^2$ for a Planck-mass black hole in 2+1
dimensions, so there is considerable room for information to be stored
in microscopic states; it is only when $GM\!\sim\!1/k^2\!\ll\!1$ that
the number of microscopic states of the black hole becomes small.

The most important question, of course, is whether the results of
this paper can be extended to 3+1 dimensions.  In the form presented
here, they clearly cannot: (3+1)-dimensional general relativity has
no Chern-Simons formulation, and there is no easy way to view the
diffeomorphisms as an ordinary group of gauge transformations.
Nevertheless, it is plausible that the underlying physical ideas
introduced here can be translated into the formalism of standard
general relativity.  In particular, the boundary term \rref{b3} has
a metric counterpart in the $\hbox{Tr}\,K$ term that must be added to
the Einstein action on a manifold with boundary.  This term is invariant
only under those transformations that take $\partial M$ to itself, so
the metric degrees of freedom that would normally be eliminated by
``diffeomorphisms'' normal to $\partial M$ ought to become dynamical,
providing new physics at the black hole horizon.  Whether this
description can be made quantitatively correct remains to be seen.

\vspace{1.5ex}
\begin{flushleft}
\large\bf Acknowledgements
\end{flushleft}

This work was supported in part by National Science Foundation grant
PHY-93-57203 and Department of Energy grant DE-FG03-91ER40674.  A
portion of the work was performed at the Isaac Newton Institute, for
whose hospitality I am grateful; while there, I received partial
support from EPSRC grant GR/J90015.

\end{document}